\documentstyle[aps,multicol,pre,epsf]{revtex}
   \begin{document}
   \def\I.#1{\it #1}
   \def\B.#1{{\bbox#1}}
   \def\C.#1{{\cal #1}}
  % \input REVTeXDraft1.tex

%   \title{{ Draft for Phys. Rev. Lett. \hfill Version of \today } \\~~\\
   \title{Laplacian Growth and Diffusion Limited Aggregation: different
   universality classes}
   \author {Felipe Barra, Benny Davidovitch, Anders Levermann and Itamar
   Procaccia}
   \address{Department of~~Chemical Physics, The Weizmann Institute of
   Science, Rehovot 76100, Israel}
   \maketitle
%%%%%%%%%%%%%%%%%%%%%%%%%%%%%%%%%%%%%%%%%%%%%%%%%%%%%%%%%%%%%%%%%%%%%%%%%%%%%%
%
% Abstract
%
%%%%%%%%%%%%%%%%%%%%%%%%%%%%%%%%%%%%%%%%%%%%%%%%%%%%%%%%%%%%%%%%%%%%%%%%%%%%%%

   \begin{abstract}
   It had been conjectured that Diffusion Limited Aggregates and Laplacian
   Growth Patterns (with small surface tension) are in the same
   universality class.
   Using iterated conformal maps we construct
   a 1-parameter family of fractal growth patterns with a continuously
   varying fractal dimension.
   This family can be used to bound the dimension of Laplacian Growth
   Patterns from below. The bound value is higher than the dimension of Diffusion
   Limited Aggregates,
   showing that the two problems belong to two different universality
   classes.
   \end{abstract}
   \pacs{PACS numbers 47.27.Gs, 47.27.Jv, 05.40.+j}
\begin{multicols}{2}\narrowtext

   Laplacian Growth Patterns are obtained when the boundary $\Gamma$ of a
   2-dimensional
   domain is grown
   at a rate proportional to the gradient of a Laplacian field $P$. Outside
   the domain
   $\nabla^2 P=0$, and each point of $\Gamma$ is advanced at a rate
   proportional
   to $\B.\nabla P$ \cite{58ST}. It is well known that without ultra-violet
   regularization such
   growth results in finite time singularities \cite{84SB}.
   In correspondence with experiments
   on viscous fingering one usually adds surface tension, or in other words
solves the above problem with the boundary condition $P=\sigma\kappa$
   where $\sigma$ is the surface tension and $\kappa$ the local curvature
   of
   $\Gamma$ \cite{86BKT}.
   The other boundary condition is that as $r\to \infty$ the flux is
   $\B.\nabla P={\rm const}\times\hat r/r$.
   Fig. 1 (left) shows a typical Laplacian Growth Pattern.
\begin{figure}
\hskip -1.0 cm
\epsfxsize=5truecm
\epsfbox{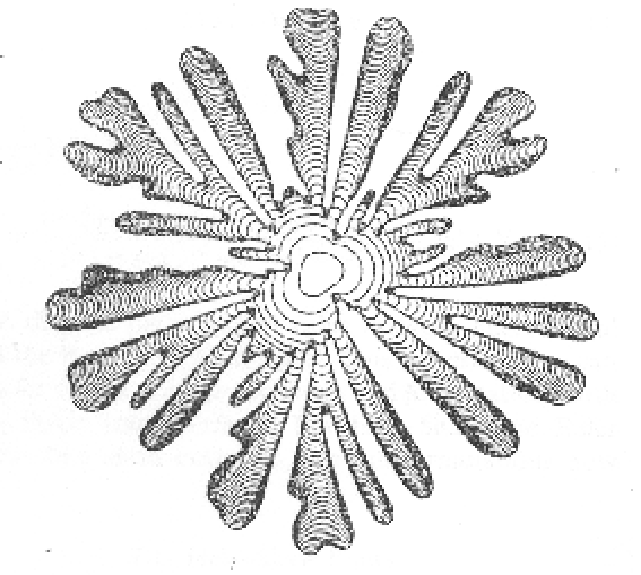}
\epsfxsize=4.5truecm
\epsfbox{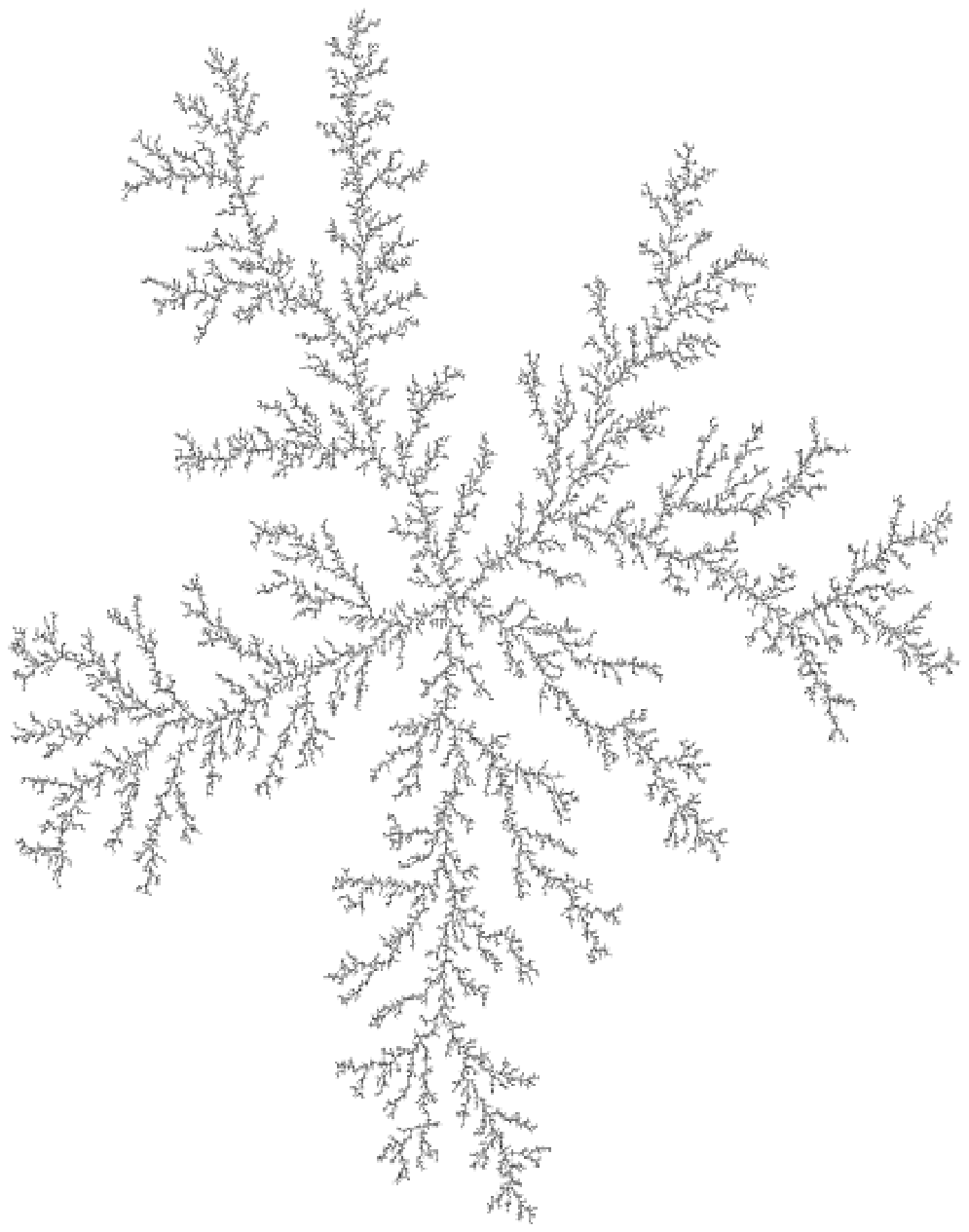}
\caption{Left: Typical Laplacian Growth pattern with surface tension, 
cf. Ref.[6].
Right: Typical DLA cluster of 100000 particles.}
\label{patterns}
\end{figure}
   Diffusion Limited Aggregation (DLA) \cite{81WS} begins with fixing one
   particle at the center of
   coordinates in 2-dimensions, and
   follows the creation of a cluster by releasing fixed size random walkers
   from infinity,
   allowing them to walk around until they hit any particle belonging to
   the
   cluster. Since the particles are released one by one
  and may take arbitrarily long time to hit the cluster, the probability field
is quasi-stationary and in the
   complement of the cluster we have again $\nabla^2 P=0$. In this case the
 boundary condition on the cluster is $P=0$, but finite time singularities
are avoided by having finite size particles. The boundary condition
   at infinity is exactly as above. A typical DLA is shown on Fig.1 (right).

   In spite of the different ultra-violet regularizations of Laplacian
   Growth and DLA, it was speculated by many authors \cite{84Pat}
   that the two problems belong to the same ``universality class",
   and it was expected that the resulting
   fractal patterns will have the same dimension. In this Letter we argue
   that
   this is not the case: there are deep differences between the two
   problems, and
   in particular Laplacian Growth Patterns have a dimension
   considerably higher than DLA. In one sentence, the differences between
   the problems stem from the fact that Laplacian Patterns are grown layer by
   layer, whereas DLA is grown particle by particle. Unfortunately,
   traditional
   techniques used to grow Laplacian Growth patterns, either numerical
   \cite{94HLS} or experimental \cite{85BGGKLMS}, fail to achieve patterns large
   enough to extract reliable dimensions
   (and see Fig.9 in \cite{94HLS} for example). The numerical algorithms
   are extremely time consuming due to the stiffness of the equations
   involved;
   experimentally it is difficult to construct large quasi
   two-dimensional (Hele-Shaw)
   cells without introducing serious deformations.

   The aim of this Letter is to provide a scheme to simulate the zero surface
tension Laplacian Growth that has a finite size regularization and thus does
not suffer from finite time singularities. We 
introduce a 1-parameter family of growth
   processes based on iterated conformal maps \cite{98HL,99DHOPSS}.
   Contrary to DLA which grows particle by particle, we will construct the
   family of growth processes to mimic Laplacian Growth, in which a layer
   is added to the boundary $\Gamma$ at each growth step, with a width
   proportional
   to the gradient of the field. Consider then
   $\Phi^{(n)}(w)$ which conformally maps the exterior of the unit circle
   $e^{i\theta}$ in the
   mathematical $w$--plane onto the complement of the (simply-connected)
   cluster of $n$ particles in the physical $z$--plane.
   The unit circle is
   mapped onto the boundary of the cluster. The map $\Phi^{(n)}(w)$ is
   made from compositions of elementary maps $\phi_{\lambda,\theta}$,
   \begin{equation}
   \Phi^{(n)}(w) = \Phi^{(n-1)}(\phi_{\lambda_{n},\theta_{n}}(w)) \ ,
   \label{recurs}
   \end{equation}
   where the elementary map $\phi_{\lambda,\theta}$ transforms the unit
   circle to a circle with a ``bump" of linear size $\sqrt{\lambda}$ around
the point $w=e^{i\theta}$. In this Letter we employ the elementary map
   \cite{98HL}
   \begin{eqnarray}
   &&\phi_{\lambda,0}(w) = \sqrt{w} \left\{ \frac{(1+
   \lambda)}{2w}(1+w)\right. \nonumber\\
   &&\left.\times \left [ 1+w+w \left( 1+\frac{1}{w^2} -\frac{2}{w}
   \frac{1-
   \lambda} {1+ \lambda} \right) ^{1/2} \right] -1 \right \} ^{1/2} \\
   &&\phi_{\lambda,\theta} (w) = e^{i \theta} \phi_{\lambda,0}(e^{-i
   \theta}
   w) \,,
   \label{eq-f}
   \end{eqnarray}
   With this choice the map $\Phi^{(n)}(w)$ adds on a
   new semi-circular bump to the image of the unit circle under
   $\Phi^{(n-1)}(w)$. The
   bumps in the $z$-plane simulate the accreted particles in
   the physical space formulation of the growth process.
   The recursive dynamics can be represented as iterations
   of the map $\phi_{\lambda_{n},\theta_{n}}(w)$,
   \begin{equation}
   \Phi^{(n)}(w) =
   \phi_{\lambda_1,\theta_{1}}\circ\phi_{\lambda_2,\theta_{2}}\circ\dots\circ
\phi_{\lambda_n,\theta_{n}}(\omega)\ . \label{comp}
   \end{equation}

 With the present technique it is also straightforward
   to determine the dimension. The conformal map $\Phi^{(n)}(\omega)$
   admits
   a Laurent expansion
   \begin{equation}
   \Phi^{(n)}(\omega) = F_1^{(n)} \omega +F_0^{(n)}
   +\frac{F_{-1}^{(n)}}{\omega}+~\cdots \ .
   \end{equation}
   The coefficient of the linear term is the Laplace radius, and was shown
   to scale like
   \begin{equation}
   F_1^{(n)}\sim S^{1/D} \ ,
   \end{equation}
   where $S$ is the area of the cluster (the sum of the actual areas of the
   bumps in the
   physical space). On the other hand $F_1^{(n)}$ is given analytically by
   \begin{equation}
   F_1^{(n)} = \prod_{k=1}^n \sqrt{(1+\lambda_k)} \ ,
   \end{equation}
   and therefore can be determined very accurately.

   Different growth processes can be constructed by proper choices of the
   itineraries $\{\theta_i\}_{i=1}^n$ \cite{00DFHP}, and rules for
   determining
   the areas of the bumps $\{\lambda_i\}_{i=1}^n$.
   In DLA growth \cite{98HL,99DHOPSS} one wants
   to have {\em fixed size} bumps in the physical space, say
   of fixed area $\lambda_0$. Then one chooses in the $n$th step
   \begin{equation}
   \lambda_{n} = \frac{\lambda_0}{|{\Phi^{(n-1)}}' (e^{i \theta_n})|^2}
   \ , \quad{\rm DLA~growth} \ .
   \label{lambdan}
   \end{equation}
   The probability to add a particle to the boundary of the DLA cluster is
   the harmonic measure, which is uniform on the circle. Thus in DLA the
   itinerary
   $\{\theta_i\}_{i=1}^n$ is random, with uniform probability for
   $\theta_i$
   in the interval $[0,2\pi]$.

   For our present purposes we want to grow a {\em layer} of particles of
   varying sizes, proportional to the gradient of the field, rather than
   {\em one} particle of fixed size. This entails three major changes.
   First, if we want to grow {\em one} particle of size proportional
   to the gradient of the field, (i.e. area proportional to
   $|{\Phi^{(n-1)}}' (e^{i \theta_n})|^{-2}$) we need to choose
   \begin{equation}
   \lambda_{n} = \frac{\lambda_0}{|{\Phi^{(n-1)}}' (e^{i \theta_n})|^4} \ ,
\quad{\rm present~models} \ . \label{lambdanew}
   \end{equation}
   Second, to grow a layer, we need to accrete many particles without
   updating the conformal map. In other words, to add a new layer of $p$
   particles
   when the cluster contains $m$ particles, we need to choose $p$
   angles on the unit circle $\{\tilde\theta_{m+k}\}_{k=1}^p$. At these
   angles we grow bumps which in the physical space are proportional
   in size to the gradient of the field around the $m$-particle cluster:
   \begin{equation}
   \lambda_{m+k} = \frac{\lambda_0}{|{\Phi^{(m)}}'
   (e^{i \tilde\theta_{m+k}})|^4} \ , \quad k=1,2\dots, p \ . \label{layer}
\end{equation}
   Lastly, and very importantly, we need to choose the itinerary
   $\{\tilde\theta_{m+k}\}_{k=1}^p$ which defines the layer. This itinerary
is chosen to achieve a uniform coverage of the unit circle {\em before}
   any growth takes
   place. The parameter that will distinguish one growth model from
   another,
   giving us a 1-parameter control, is the {\em degree of coverage}.
   In other words we introduce the parameter
   \begin{equation}
   {\cal C}=\frac{1}{\pi}\sum_{k+1}^p \sqrt{\lambda_{m+k}} \ .
   \end{equation}
   This parameter is the fraction of the unit circle which is covered in
   each layer,
   with the limit of Laplacian Growth obtained with ${\cal C}=1$.
  It turns out to be rather time consuming to grow fractal patterns
  with ${\cal C}$ close to unity. But we will show below that this is
  hardly necessary; already for ${\cal C}$ of the order of 1/2 we will
  find patterns whose fractal dimension significantly exceeds that
  of DLA, offering a clear lower bound on the dimension of Laplacian
  Growth patterns.

   Once a layer with coverage ${\cal C}$ had been grown, the field is
   updated. To do this, we define a series $\{\theta_k\}^p_{k=1}$ according
   to
   \begin{equation}
   \Phi^{(m)}(e^{i\tilde\theta_{m+k}})\equiv
   \Phi^{(m+k-1)}(e^{i\theta_{m+k}})\ .
   \end{equation}
   Next we define the conformal map used in the next layer growth according
   to
   \begin{equation}
   \Phi^{(m+p)}(\omega)\equiv \Phi^{(m)}\circ\phi_{\theta_{m+1},\lambda_{m+1}}\circ
   \dots\circ
   \phi_{\theta_{m+p},\lambda_{m+p}}(\omega) \ .
   \end{equation}
   It is important to notice that on the face of it this conformal map
   appears very similar to the one obtained in DLA, Eqs.(\ref{recurs}),
   (\ref{comp}). But this is deceptive; the distribution of $\theta$ values is different,
we do not update the map after each particle, and the growth rule is different.

   We can achieve a uniform coverage ${\cal C}$ using various itineraries.
   One way is to construct the ``golden mean trajectory"
   $\tilde\theta_{m+k+1}=\tilde\theta_{m+k}+2\pi \rho$
   where $\rho=(\sqrt{5}-1)/2$. At each step we check whether the newly
   grown
   bump may overlap a previous one in the layer. If it does, this growth
   step is skipped and the orbit continues until a fraction ${\cal C}$ is
   covered. Another method is random choices of $\tilde\theta_{m+k}$ with
   the same rule of skipping overlaps. We have tried several other
   itineraries. Of course, to be an acceptable model of Laplacian
growth the resulting cluster should be invariant to the itinerary. This
invariance is demonstrated below. The central thesis of this work is that 
the dimension of the resulting
   growth patterns is dependent on ${\cal C}$ only, and not on the
   itinerary chosen to achieve it. Numerically it is more efficient to use the golden
mean itinerary since it avoids as much as possible previously visited
   regions. In order to achieve comparable growth rates for different
   layers we inflated $\lambda_0$ in Eq.(\ref{layer}) according to
   $\lambda_0 \to m \lambda_0$ in the layer composed of $p$ particles
   $\{m+k\}_{k=1}^p$.
In Fig. 2 we show $F_1$ of clusters grown by choosing 3 different
   itineraries to produce the layers and for two values of ${\cal C}$.
\begin{figure}
\epsfxsize=7truecm
\epsfbox{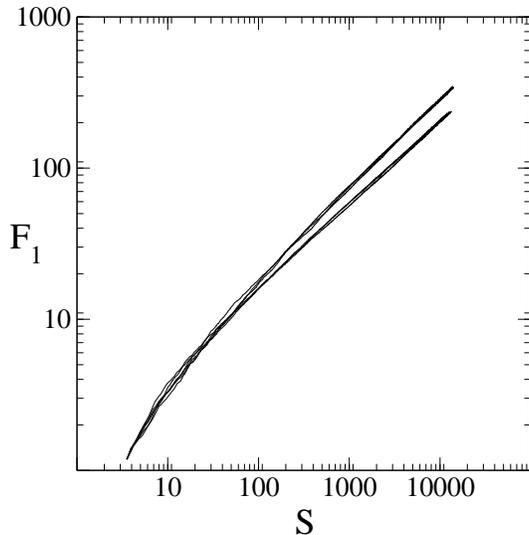}
\caption{Log-log plots of $F_1$ vs. $S$ of six individual clusters, using 3 
differnt itineraries for layer construction, with two values of ${\cal C}$.  
${\cal C} =0.3$ (upper group) and ${\cal C} =0.5$ (lower
group). Here we use the golden-mean, random and the period doubling
itineraries (see Ref.[10]).} 
\label{Fig2}
\end{figure}
   We conclude that the dimension (determined by the asymptotic behavior of
$F_1$ vs. $S$) does not depend on the itinerary used to form the layers
   but on ${\cal C}$ only.

   In Fig.3 we show three fractal patterns grown with this method, with
   three different values of ${\cal C}$. Even a cursory observation should
   convince the reader that the dimension of these patterns grows upon
   increasing ${\cal C}$.
\begin{figure}
\hskip -0.5cm
\epsfxsize=5truecm
\epsfysize=5truecm
\epsfbox{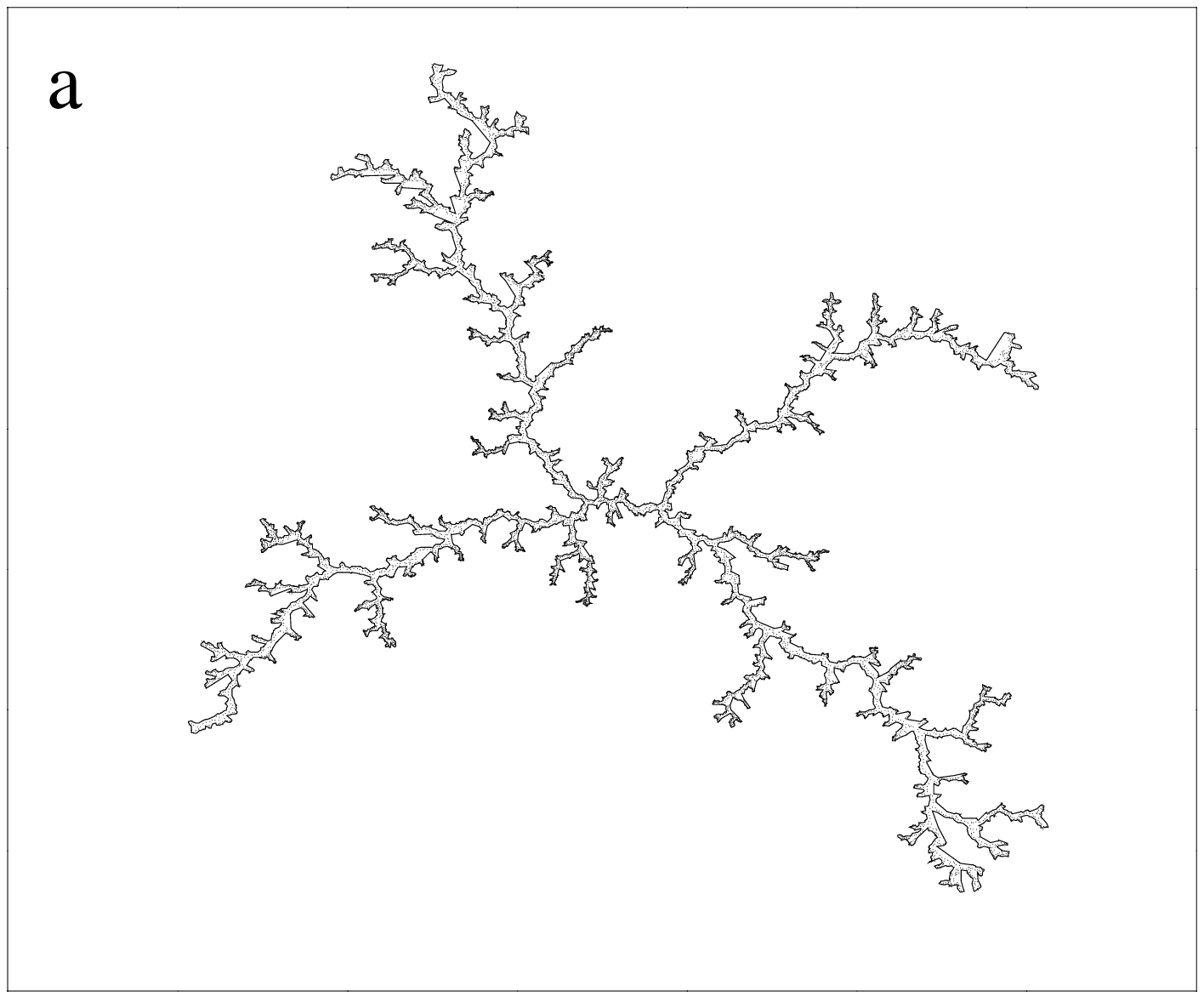}
\epsfxsize=5truecm
\epsfysize=5truecm
\epsfbox{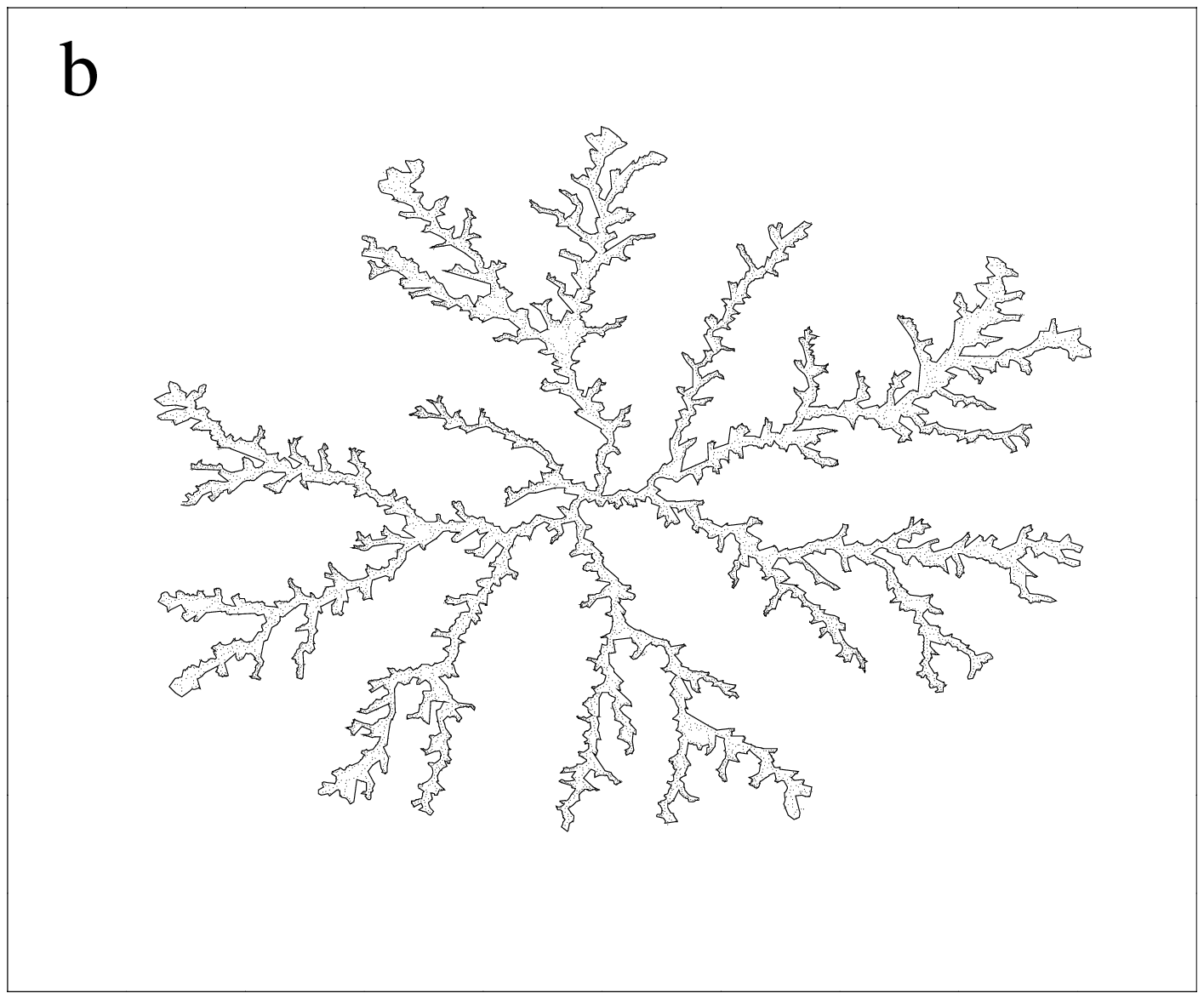}
\vskip 0.2cm
\hskip 2cm
\epsfxsize=5truecm
\epsfysize=5truecm
\epsfbox{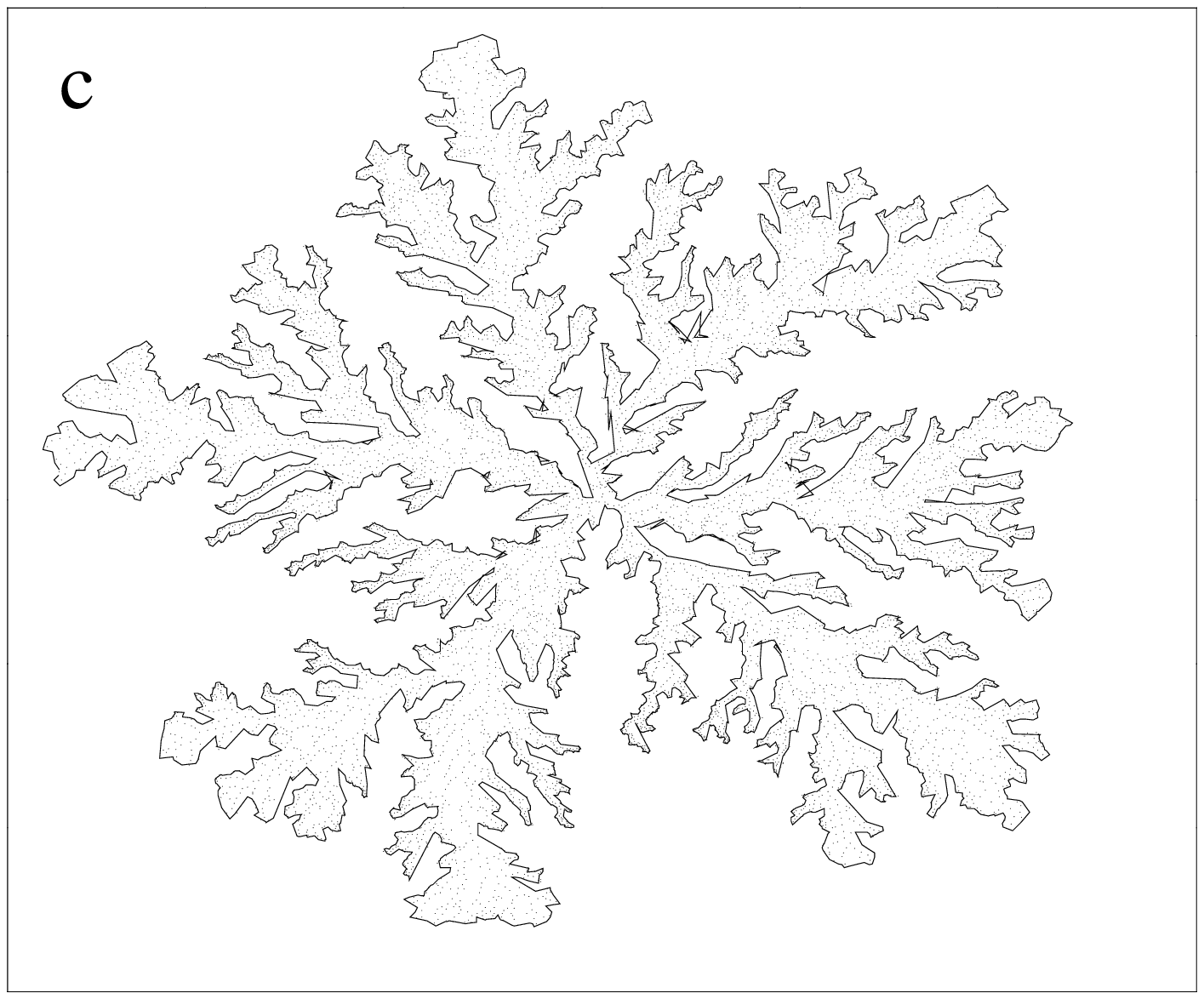}
\caption{Patterns grown with 3 different values of ${\cal C}$ by using the
golden-mean itinerary: a) ${\cal C} =0.1$, b) ${\cal C} =0.3$, 
c) ${\cal C} =0.5$.}
\label{Fig3}
\end{figure}
In order to calculate the dimension we averaged $F_1$ of many
   clusters produced by the golden mean itinerary, each with another
random initial angle in each layer.
   Plots of the averages $ \langle F_1 \rangle$
   for 3 values of ${\cal C}$ are presented in Fig. 4.
\begin{figure}
\epsfxsize=7truecm
\epsfbox{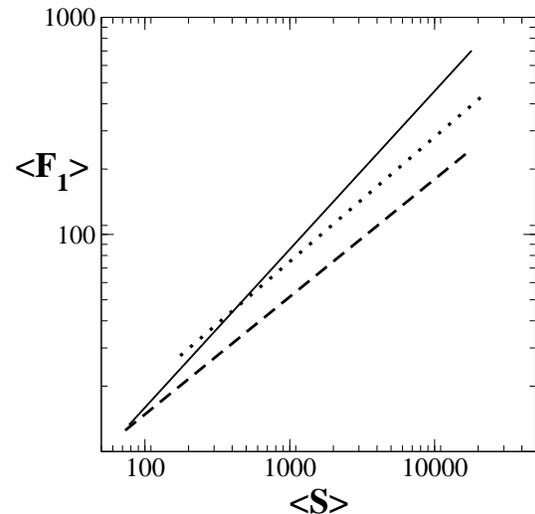}
\caption{Linear regressions of log-log plots of 
$ \langle F_1 \rangle$ vs. $S$ for 3 values of 
${\cal C}$: 0.1 (solid line), 0.3 (dotted) and 0.6 (dashed). The slopes of
the curves imply dimensions D=1.37, D=1.75 and D=1.85 respectively. 
The averages are taken over at least 20 clusters.}
\label{Fig4}
\end{figure}
   We conclude that the dimension of the growth pattern increases
   monotonically with ${\cal C}$, with $D \approx 1.85$ when ${\cal
   C}=0.6$.

   The main point of this analysis is that the dimension of Laplacian
   growth patterns is bounded from below by the supremum on the dimensions
   obtained in this family of models. First,
   Laplacian Growth calls for ${\cal C}=1$. Second, in Laplacian Growth
   the boundary condition is $P=\sigma\kappa$, suppressing growth at the
   tips (and relatively favoring growth in the fjords) compared to growth
   with the boundary condition $P=0$. Accordingly, on the basis of the
   results
   shown in Fig.4, we propose that the dimension of Laplacian Growth
   patterns
   exceeds 1.85, putting it distinctively away from the dimension of DLA
   which is about 1.71 \cite{83Mea}.

  In hindsight, it is difficult to understand how the consensus formed in
favor of DLA and Laplacian Growth being in the same universality class.
Superficially one could say that in DLA the update of the harmonic measure
after each particle is not so crucial, since the effect of such an update is relatively {\em local}
 \cite{94Hal}. Thus it may just work that a full layer of particles would be
added to the cluster before major interaction between different growth
events takes place. However this view is completely wrong. An incoming random
walker lands on top of a previously attached one {\em very often}. 
To see this, consider how many 
   angels $\{ \theta_j \}$ can be chosen {\em randomly} on the unit circle
   before the first overlap between bumps (of linear sizes
   $\epsilon_j = \sqrt{\lambda_n (e^{i \theta_j})}$).
To get the order of magnitude take $\epsilon_j = \epsilon=\langle\sqrt{\lambda_n }\rangle$.
   The average number of times that we can choose randomly an angle before the first overlap
 is ${\cal N}(\epsilon) \sim \frac{1}{\sqrt{\epsilon}}$. The Length
  of the unit circle that is covered at that time by the already chosen bumps
   is ${\cal L}(\epsilon) = \epsilon {\cal N}(\epsilon) \sim \sqrt{\epsilon}$.
It was shown in \cite{99DHOPSS} that for DLA $\langle\lambda_n\rangle \sim \frac{1}{n}$,
so that $\epsilon \sim \frac{1}{\sqrt{n}}$, implying 
   ${\cal N}(n) \sim n^{1/4}$.
   Notice that this result means in particular that for a DLA cluster of 1
   million particles
   only less than 50 random walkers can be attached before two of them will
   arrive at the same site! Moreover, ${\cal L}(n) \sim \frac{1}{n^{1/4}} 
\to 0$ for $n \to \infty$,
   which means that as the DLA cluster grows, our coverage parameter ${\cal C}$
goes to zero, rather than to unity where Laplacian Growth is. Taking spatial fluctuations of $\lambda_n$ into
   account may change the exact exponents but not the qualitative result.
   This argument clarifies the profound difference between growing
   a whole layer simultaneously and particle-by-particle. Note however that DLA
is NOT the ${\cal C}\to 0$ limit of our 1-parameter family due to the difference between
Eqs.(\ref{lambdan}) and (\ref{lambdanew}).

   The results of this study underline once more the delicacy of the issues
   involved. Fractal patterns depend sensitively on the details of
   the growth rules. Even though the analytic presentation seems very
   similar,
   to the degree that many researchers were led to believe in wide
   universality classes, we showed here that one must be much more
   cautious.
   By lifting the models into {\em families} of growth patterns depending
   on a parameter we could demonstrate strong variability of the fractal
   dimension. Here we constructed the family to bound from below Laplacian
   Growth patterns.A similar family can be constructed to bound DLA from above.
   This and other aspects of this method will be reported elsewhere.
   %%%%%%%%%%%%%%%%%%%%%%%%%%%%%%%%%%%%%%%%%%%%%%%%%%%%%%%%%%%%%%%%%%%%%%%%%%%%%%

   %
   % Acknowledgments
   %
   %%%%%%%%%%%%%%%%%%%%%%%%%%%%%%%%%%%%%%%%%%%%%%%%%%%%%%%%%%%%%%%%%%%%%%%%%%%%%%

   \acknowledgments
   This work has been supported in part by the
   European Commission under the TMR program and the Naftali and Anna
   Backenroth-Bronicki Fund for Research in Chaos and Complexity.

   %%%%%%%%%%%%%%%%%%%%%%
   
\end{multicols}
\end{document}